\documentclass[sigconf,nonacm,screen]{acmart}

\usepackage{xspace}
\usepackage{multicol}
\usepackage{tabularx}

\newcommand{\eg}{e.g.,\xspace}

\newcommand{\Parabreak}{1.5ex}
\newcommand{\Paragraph}[1]{\vspace{\Parabreak}\noindent\textbf{#1}}
\newcommand{\swsc}{software supply chain\xspace}
\newcommand{\swscs}{software supply chain security\xspace}

\usepackage{array}
\newcolumntype{x}{>{\centering\arraybackslash\hspace{0pt}}c@{\hspace{0.25\tabcolsep}}}
\newcolumntype{y}{>{\centering\arraybackslash\hspace{0pt}}c@{\hspace{1.5\tabcolsep}}}
\newcolumntype{k}[1]{>{\centering\arraybackslash\hspace{0pt}}m{#1}@{\hspace{0.25\tabcolsep}}}
\newcolumntype{z}[1]{>{\centering\arraybackslash\hspace{0pt}}m{#1}@{\hspace{1.5\tabcolsep}}}

\newcolumntype{R}[2]{%
    >{\adjustbox{angle=#1,lap=\width-(#2)}\bgroup}%
    l%
    <{\egroup}%
}
\usepackage{tikz}
\usetikzlibrary{arrows.meta}
\newcommand*\fullcirc[1][0.75ex]{\tikz\fill (0,0) circle (#1);} 

\setcopyright{none} % No copyright notice required for submissions
\acmConference[Anonymous Submission \#9 to ACM SCORED '22]{ACM Workshop on Software Supply Chain Offensive Research and Ecosystem Defenses}{Due August 5, 2022}{Los Angeles, CA (and virtual)}
\acmYear{2022}

\begin{document}
\title{What is Software Supply Chain Security?} % TODO: replace with your title

\author{Marcela S. Melara}
\affiliation{%
  \institution{Intel Labs}
  \city{Hillsboro}
  \state{OR}
  \country{}}
\email{marcela.melara@intel.com}

\author{Mic Bowman}
\affiliation{%
  \institution{Intel Labs}
  \city{Hillsboro}
  \state{OR}
  \country{}}
\email{mic.bowman@intel.com}

\begin{abstract}
The \swsc involves a multitude of tools and processes that
enable software developers to write, build, and ship applications.
Recently, security compromises of tools or processes has led to
a surge in proposals to address these issues.
However, these proposals commonly overemphasize specific solutions
or conflate goals, resulting in unexpected consequences, or unclear 
positioning and usage.

In this paper, we make the case that developing practical solutions
is not possible until the community has a \emph{holistic} 
view of the security problem; this view must include both
the technical and procedural aspects.
To this end, we examine three use cases to identify 
common security \emph{goals}, and
present a goal-oriented taxonomy of existing solutions demonstrating
a holistic overview of \swscs.
\end{abstract}

% TODO: replace this section with code generated by the tool at https://dl.acm.org/ccs.cfm


 % TODO: replace with your keywords

\maketitle

\section{Introduction}
Applications today are built through a complex \swsc that involves
a multitude of technical and procedural aspects: first-party source code,
third-party dependencies, source control processes, build and packaging tools, package management
services.
Amid high-profile attacks~\cite{leftpad, logfourj, solarwinds-fireeye, gh-actions-vuln, pypi-typosquatting} 
that take advantage of weaknesses and user error in the \swsc, 
software developers are now focusing on addressing security.

As different parties rely on various components and processes, 
their domain-specific security requirements give rise to a plethora of security use cases.
For instance, a software engineer may want to know that an open source library they 
downloaded from a package management service, such as apt or the Python Package Index (PyPI), 
to integrate into their product was created by the expected developer from the source files and dependencies 
listed in the library's code repository. 

\Paragraph{Status Quo.}
We observe two key challenges as different communities seek to address this problem.
First, inconsistent or incomplete threat modeling leads to \emph{overemphasizing} one particular approach to
\swscs without considering compounding factors that impact risk. 

For example, the recent push by PyPI to require two-factor authentication
for updates to popular python packages~\cite{pypi-2fa} was intended to protect developers and users
against malicious software updates due to account compromise.
However, this emphasis on hardening user authentication did not take into account 
the procedural onus on developers and possible consequences: one developer's decision to bypass the 
requirements~\cite{atomicwrites-takedown} impacted hundreds of dependent packages~\cite{atomicwrites-deps}.

Second, despite the vast range of security requirements, business processes and existing tooling,
many projects attempt to provide a single solution that \emph{conflates} multiple objectives.
One example is the Security Levels for Software Artifacts (SLSA)~\cite{slsa} framework for quantifiable
metrics that indicate different levels of software artifact integrity. 

To enable users to achieve these metrics, SLSA provides a data format for capturing 
information about the build process of a software package. 
However, in doing so, SLSA shares similarities with software bill 
of materials (SBOM) proposals (\eg~\cite{spdx}) 
that aim to capture information about the constituents of a software package.
Thus, this approach has led to uncertainty about the positioning, 
functionality and usage of SLSA~\cite{slsa-positioning,slsa-positioning2}.

%A major challenge in \swscs is the need to capture information about analog processes, such as human code 
%reviews or manual tool configuration, in digital form to enable programmable/automated verification of 
%this information by an interested party. 
%This challenge often leads to solutions that conflate the task of collecting such rich and highly subjective
%information from analog sources, with the task of developing a suitable digital format that allows a computer 
%program to easily parse and evaluate this data based on a human-defined policy, all while ensuring the integrity
%of this data.

\Paragraph{Our position.} Developing practical solutions for \swscs is not
possible until security experts and software developers take a \emph{holistic}
view of the security problem space. This view must include both the technical and 
procedural aspects, allowing them to articulate concrete security goals.

We refer to any software- or hardware-based tools, online services, and data used in the 
creation of a software artifact as \emph{technical} aspects.
By \emph{procedural} aspects, we mean the manner in which an individual or organization
performs all the tasks necessary to create a software artifact. These tasks may be automated through 
the use of technology, such as a CI/CD service for unit testing, or manual, such as writing a
configuration file for the CI/CD service.

Prior work~\cite{aeva-survey} provides a set of high-level common categories and abstractions 
used in many \swscs projects today.
However, this work focuses primarily on categorizing a list of existing tools, 
and does not disentangle the programmer or business practices affected by the usage of these tools.

\Paragraph{Our contributions.} 
We demonstrate a holistic view of \swscs by studying three representative use cases and 
deriving a set of common security \emph{goals} in three main problem areas: 
trust establishment, the development of resilient tools, and resilient processes. 
These goals then allow us to build a goal-oriented taxonomy for a sample of existing solutions that
provides a first overview of overlap and gaps in current efforts.
\section{Identifying Holistic Security Goals}

To derive a set of holistic \swscs goals, we study three use cases that highlight different
security concerns, processes, and technical requirements. 

%For each case study, we ask two analysis questions:
%\begin{itemize}
%\item What are concrete security goals in each use case?
%\item Do these goals involve technical or procedural aspects of the \swsc?
%\end{itemize}
%
%These questions allow us to identify high-level problem areas within \swscs,
%and express a set of common security goals within each area that
%explicitly capture the technical and procedural aspects. 

\subsection{Case Studies}
\label{sec:case-studies}

\Paragraph{Use Case 1.} 
A software engineer is writing a Python package, which in turn includes a few open-source native and
Python libraries. She knows that the source code for these libraries is hosted on %a number of
repositories on GitHub. However, because she downloads the built libraries from package manager services
apt and PyPI, she wants to be sure that these libraries were created by the expected developers 
from the source files and dependencies listed in the packages' corresponding code repositories.

In this scenario, the software engineer has two primary security concerns 
regarding the libraries she downloads: an attacker uploaded a malicious version of a library 
replacing the version created by the legitimate author, and unexpected code was included in a library
possibly introducing security vulnerabilities.

As such, the concrete security goals are:
\begin{enumerate}
\item Verify that the downloaded library was indeed uploaded by the expected developer.
\item Verify that the open-source code repository was not tampered with by an untrusted entity.
\item Verify that the expected source materials were used to create the downloaded library.
%\item Can we check if the build pipeline for the resulting library was tampered with?
\end{enumerate}

We observe a number of technical and procedural aspects of the \swsc within these security goals.
The identity of a software developer and software components of an artifact in goals (1) and (3) are technical
aspects, whereas source control, code contribution and build in goals (2) and (3)
are procedural aspects.

What these three security goals have in common is that they seek to \emph{establish trust}
in the \swsc by verifying information about the participants or processes involved in the creation of an artifact.

%\ie a software developer, which is a technical aspect of the \swsc.~\cite{code-signing}.\footnote{Establishing the binding between a real-world individual and a digital identity is a separate problem with complex technical, procedural and \emph{social} factors; as such it is beyond the scope of this paper.}
%The second security question relates to the resilience of the source control \emph{process}, 
%such as establishing authorized code contributors, the code review process (if any). Today's source control
%\emph{services} provide features to aid repository owners in hardening this process (\eg GitHub~\cite{gh-access}), 
%though the proper use of these tools is still dependent on user configuration.
%The third security question has a technical \emph{and} a procedural component: verifying the \emph{components}
%of an artifact requires information about these components; \emph{capturing} this data about source materials can be 
%done as part of the source control and/or build process of the library.
%A number of \swscs proposals seek to enable the collection and verification of such data (\eg~\cite{in-toto,slsa,spdx}).

%AQ4: (1) signed commits, code signing (tech) (2) signed SBOM (proc/tech), code reviews (proc, be) (1a) usage: 2FA for package manager (be) (2a) CI/CD security best practices (proc/be), resilient build env: compcert, TEE (proc, tech) (2b) dependency trackers (proc, tech), checking dep code sigs (proc), controlling merges/code reviews (proc, be)

\Paragraph{Use Case 2.}
A sysadmin for a bank is set to deploy the new cloud-based banking application. Even though this banking application was developed in-house,
the application was built and tested using a third-party CI/CD service, such as Travis CI.
Before deployment, the sysadmin needs to ensure that the
application was built using the expected build process, and does not contain code vulnerabilities
that may reveal sensitive customer data.

The sysadmin in this use case is concerned with the threat of a vulnerable build process
that may output insecure code, as well as vulnerable code that could leak data, either through programmer
error or compromised dependencies.

Thus, we identify five security goals:
\begin{enumerate}
\item Verify that the build process followed the expected steps.
\item Ensure that the build tools do not contain vulnerabilities.
\item Ensure that the first-party application does not contain data leak vulnerabilities.
\item Identify all third-party dependencies.
\item Ensure that the third-party dependencies do not contain vulnerabilities.
\end{enumerate}

Goals (1) and (4) have procedural aspects related to the build process as a whole and dependency
tracking.
The technical aspects in these goals include the tools and services used throughout the build process,
as well as the properties of first- and third-party binary artifacts in goals (2), (3) and (5). 

At their core, these security goals are concerned with establishing the trustworthiness of
processes and software artifacts, and the use of resilient tools and hardened processes.

%AQ3: (1a) be: institutional trust, tech: digital identity of builder. (1b) proc: config, tech: build env (2a) proc: build config, tech: language/implementation, (2b) proc: dependency management
%
%AQ4: (1a) SLSA/signed builds (tech, be) (1b) CI/CD best practices (proc/be), resilient build env: TEE, compcert (proc, tech) (2a) static analysis, fuzzing (proc, tech) (2b) vuln scanning, static analysis/fuzzing (proc, tech)

\Paragraph{Use Case 3.}
Three hospitals seek to collaborate using federated machine learning to train a shared model for medical imaging classification. They do so by aggregating results from a locally-run learning 
algorithm based on confidential patient data, which is shipped to them as a self-contained Docker image.
Before joining the collaborative effort, the participants want to ensure they
will all be running the same container for the agreed-upon algorithm.

In this use case, the participants are concerned that the machine learning model
will be corrupted, or that confidential data will be leaked, as a result of a container that has been tampered with.

As such, the participants have three security goals:
\begin{enumerate}
\item Verify that the build process was not tampered with.
\item Verify that the container includes the expected code.
\item Ensure that the algorithm code behaves as expected.
\end{enumerate}

Goal (1) in this scenario touches on a procedural aspect of the \swsc, 
the container packaging process; the integrity of an artifact and 
ensuring \emph{correct} code behavior in goals (2) and (3) constitute technical aspects.

Although the specific procedural and technical aspects in this use case
are different than in the previous use cases, ultimately, these security 
goals aim to establish trust in a process
and the resulting software artifacts.

%AQ3: (1a) proc: config, tech: container technology (1b) be: author, tech: digital identity (2a) tech: language/implementation, guest OS (2b) be: author, tech: digital identity
%
%AQ4: (1a) CI/CD and docker security best practices (proc, be), SBOM (proc, tech), signed builds/SLSA (proc, tech) (1b) hash, TEE (proc, tech) (2a) static analysis/fuzzing (proc, tech) (2b) code signatures, signed build/SLSA, TEE, hash (tech)

\subsection{Security Goals}
\label{sec:goals}
Our analysis in~\S\ref{sec:case-studies} reveals three overarching areas
that \swsc seeks to address: (1) trust establishment, (2) resilient tools, 
and (3) resilient processes. Based on the concrete goals for each use case,
we derive common \swscs goals within each area.

\Paragraph{(1) Trust Establishment Goals:}
\begin{itemize}
\item Verify the identity of a participant
\item Verify the components of an artifact
\item Verify properties or behavior of an artifact
\item Verify properties of a process
\end{itemize}

\Paragraph{(2) Resilient Tools Goals:}
\begin{itemize}
\item Identify vulnerabilities in an existing tool
\item Deploy high-assurance tools
\end{itemize}

\Paragraph{(3) Resilient Process Goals:}
\begin{itemize}
\item Implement a process that reduces an attack surface
\item Automate an error-prone manual process
\end{itemize}

%\Paragraph{Scope.}
%
%In addition, social aspects of security, such as developing incentives to adopt novel
%tools or processes is beyond the scope of this paper.

\begin{table*}[t!]
    \centering
    \footnotesize
    \caption{Solution requirements to achieve \swscs trust establishment goals.}
    \label{tab:trust}
\begin{tabularx}{\textwidth}{c xk{1.7cm}k{1.2cm}z{1.5cm} k{0.9cm}z{1.8cm} z{1.5cm} k{1.8cm}k{1.5cm}}
  \toprule
  \textbf{Example} & \multicolumn{4}{c}{\textbf{Trust Data Capture}} & \multicolumn{2}{c}{\textbf{Policy Definition}} & \textbf{Data Distribution} & \multicolumn{2}{c}{\textbf{Data Verification}} \\
  \midrule
   & \textbf{Identity} & \textbf{Artifact Components} & \textbf{Artifact Behavior} & \textbf{Process Properties} & \textbf{Policy Format} & \textbf{Policy Generation} & & \textbf{Trust Data Authentication} & \textbf{Trust Data Validation} \\
  \midrule
  SPIFFE/SPIRE~\cite{spiffe} & \fullcirc & & & & & & \fullcirc & & \fullcirc \\ 
  SPDX~\cite{spdx} & & \fullcirc & & & & & & & \fullcirc \\
  in-toto~\cite{in-toto} & & \fullcirc & & \fullcirc & & & & \fullcirc & \fullcirc \\
  CDI~\cite{cdi} & & & \fullcirc & \fullcirc & & & & \fullcirc \\
  SLSA~\cite{slsa,slsa-vsa} & & \fullcirc & & \fullcirc & & & & \fullcirc & \fullcirc \\
  OPA~\cite{opa} & & & & & \fullcirc & & & & \fullcirc \\ 
  Nuclei~\cite{nuclei} & & & & & \fullcirc & \fullcirc & & & \\
  Sigstore~\cite{sigstore} & & & & & & & \fullcirc & \fullcirc & \\
  \bottomrule
\end{tabularx}
\end{table*}

\section{Current Solutions: A Goal-Oriented Taxonomy}
The goals articulated in~\ref{sec:goals} allow us to identify specific requirements
and tools needed to achieve these goals.
Specifically, achieving the resilient tool goals requires techniques that aid 
\swsc \emph{tool} developers in detecting vulnerabilities, or enable the creation of 
creation of high-assurance tools, such as formal verification, language-based techniques,
or hardware-based hardening.

Addressing the resilient process goals requires instituting processes that reduce specific 
risks either through new dedicated processes or security features provided by \swsc services.
To automate error-prone manual processes, dedicated tooling that handles a specific process on
behalf of the developer is needed.

Accomplishing all trust establishment goals requires capturing information 
about identities, artifacts and \swsc processes in a verifiable representation, as well as
enabling verifiers to express and generate trust policies, and building tools that can disseminate 
or evaluate this information based on a given policy.

Given these specific requirements, we provide a holistic, goal-oriented classification of well-known examples
in today's solution space. As such, our taxonomy does not cover an exhaustive list of existing 
\swsc tools and processes.
Tables~\ref{tab:trust},~\ref{tab:tooling}, and~\ref{tab:processes} summarize our
goal-oriented taxonomy for the trust establishment, resilient tools, and resilient process goals,
respectively. 

\Paragraph{Taxonomy Insights.}
Our taxonomy reveals opportunities for growth to address tool and process resilience.
Within the high-assurance tools space in particular, we believe that building more formally
verified tools, as well as language-based and hardware-based techniques 
(\eg trusted execution~\cite{sgx,amd-sev}), can have significant impact.
For process resilience, we see a need for developing both new processes that reduce risk,
such as code security reviews, and tools to automate these processes to ease
adoption.

We also observe a trend towards conflation in the trust establishment solution space
regarding trust data capture and verification. 
This trend may be due to disparate data formats that capture different trust data. 
To address this conflation, we see a potential 
for more interoperable data formats and general verification solutions.

\begin{table}[t]
    \centering
    \footnotesize
    \captionsetup{width=.9\columnwidth}
    \caption{Solution requirements to achieve \swscs resilient tool goals.}
    \label{tab:tooling}
\begin{tabularx}{\columnwidth}{z{2.1cm} z{1.6cm} k{1.4cm}k{1.4cm}k{1.4cm}}
  \toprule
  \textbf{Example} & \textbf{Tool Vulnerability Detection} & \multicolumn{3}{c}{\textbf{High-Assurance Tools}} \\
  \midrule
   & & \textbf{Formal Verification} & \textbf{Language-Based Hardening} & \textbf{Hardware-Based Hardening} \\
  \midrule
  IFuzzer~\cite{ifuzzer}  & \fullcirc & & & \\
  Prog-Fuzz~\cite{prog-fuzz} & \fullcirc & & & \\
  GWChecker~\cite{gwchecker} & \fullcirc & & & \\
  CompCert~\cite{comp-compcert,compcertm} & & \fullcirc & & \\
  rustc~\cite{rustc} & & & \fullcirc & \\
  VM isolation~\cite{sec-ci} & & & & \fullcirc \\
  TEE~\cite{cdi} & & & & \fullcirc \\
  \bottomrule
\end{tabularx}
\end{table}

\begin{table}[t]
    \centering
    \footnotesize
    \captionsetup{width=.9\columnwidth}
    \caption{Solution requirements to achieve \swscs resilient process goals.}
    \label{tab:processes}
\begin{tabularx}{\columnwidth}{c k{1.5cm}z{1.5cm} k{1.5cm}}
  \toprule
  \textbf{Example} & \multicolumn{2}{c}{\textbf{Risk Reduction Processes}} & \textbf{Process Automation} \\
  \midrule
   & \textbf{Dedicated Processes} & \textbf{Service Security Features} & \\
  \midrule
  SDL practices~\cite{ms-sdl} & \fullcirc & & \\
  Reproducible Builds~\cite{reproducible-builds, rebuilderd} & \fullcirc & & \fullcirc \\
  GitHub roles~\cite{gh-roles} & & \fullcirc & \\
  PyPI 2FA~\cite{pypi-2fa} & & \fullcirc & \\
  dependabot~\cite{dependabot} & & \fullcirc & \fullcirc \\
  DataFlowSanitizer~\cite{df-san} & & & \fullcirc \\
  CodeGuru~\cite{codeguru} & & & \fullcirc \\
  \bottomrule
\end{tabularx}
\end{table}
\section{Conclusion}
Understanding the complex set of software development technologies as well as engineering practices 
that impact the security of a \swsc, will enable the community to more effectively address classes of use 
cases and identify the gaps not addressed by current proposals.
Ultimately, our goal is to spur deeper analysis of real use cases and existing proposals, 
and to encourage consensus and collaboration within the \swscs community.

\bibliographystyle{ACM-Reference-Format}
\bibliography{references}

\end{document}